\documentclass{aa}  
\usepackage{siunitx}
\usepackage{graphicx}
\usepackage{txfonts}
\usepackage{lipsum}
\usepackage{subcaption}         
\usepackage{lscape}             
\usepackage{placeins}           
\usepackage{natbib}   
\usepackage{xcolor}
\usepackage[hidelinks,colorlinks=true,linkcolor=blue,citecolor=blue]{hyperref}

\defcitealias{2010ApJ...724..878T}{T10}
\defcitealias{2024ApJS..272...39W}{WH24}
\defcitealias{2021ApJ...920...13M}{M21}

\begin{document}
\raggedbottom
   \title{Testing general relativity with galaxy cluster clustering from the DESI Legacy Imaging Surveys}

%
%
%

   \author{
Eleni Tsaprazi\inst{1}\corrauth{eleni.tsaprazi@obspm.fr}\fnmsep\thanks{Observatoire de Paris - PSL Fellow}
\and Federico Marulli\inst{2,3,4}
\and Pier-Stefano Corasaniti\inst{1,5}
\and Amandine~M.~C.~Le~Brun\inst{1} \and Yann Rasera\inst{1,6,7}
}

   \institute{LUX, Observatoire de Paris, Université PSL, Sorbonne Université, CNRS, 92190 Meudon, France
   \and Dipartimento di Fisica e Astronomia “Augusto Righi” - Alma Mater Studiorum Universita di Bologna, via Piero Gobetti 93/2, I-40129 Bologna, Italy \and INAF - Osservatorio di Astrofisica e Scienza dello Spazio di Bologna, via Piero Gobetti 93/3, I-40129, Bologna, Italy \and INFN - Sezione di Bologna, viale Berti Pichat 6/2, I-40127 Bologna, Italy \and Institut d’Astrophysique de Paris, CNRS and Sorbonne Université, UMR 7095, 98 bis Boulevard Arago, F-75014 Paris, France \and Université Paris Cité, F-75006 Paris \and Institut Universitaire de France}

   \date{Received ???}


\abstract{
Deviations from general relativity on cosmological scales have been proposed as possible explanations for dark energy and dark matter. Galaxy cluster clustering offers an independent probe of gravitational growth, benefiting from theoretically motivated halo-bias constraints and complementing cluster-abundance measurements. We analyse a subsample of the largest publicly available spectroscopic galaxy cluster catalogue to date, constructed from the DESI Legacy Imaging Surveys. Following our selection cuts, the sample comprises \num{85818} clusters at $0.1\leq z \leq0.4$ with $M_{\rm 500c}\gtrsim 6\times10^{13}\,M_\odot$. To enable a robust clustering analysis of this sample, we develop and validate methods for constructing random catalogues and calibrating the effective halo bias. We model the redshift-space two-point correlation function monopole and quadrupole over separations of $10$--$150\,h^{-1}\mathrm{Mpc}$ to jointly constrain gravitational growth and Alcock--Paczynski (AP) distortions, fixing the matter power-spectrum shape to that of a fiducial cosmology. We infer the product $f\sigma_8$ of the linear growth rate $f$ and the matter power spectrum normalisation $\sigma_8$, obtaining $f\sigma_8=0.45\pm0.08$ at an effective redshift of $z_{\rm eff}=0.28$, changing to $0.52\pm0.03$ when excluding baryon acoustic oscillation scales with tight AP priors. Both results are consistent with general relativity within the adopted fiducial cosmology.}


   \maketitle

\section{Introduction}

With the first cosmological data sets from late Stage-III and Stage-IV surveys becoming available \citep{2025PhRvD.112h3515A,2026arXiv260114559D}, and further improvements expected from proposed Stage-V facilities \citep{2024arXiv240305398M,2025arXiv250307923B}, we are entering a period in which tests of gravity can be performed with substantially increased precision. In this landscape, galaxy clusters are particularly interesting laboratories since they provide access to observables that are complementary to more traditional probes and can be sensitive to the transition between screened and unscreened gravitational regimes. Cluster clustering, in particular, probes the high-mass end of structure formation, provides cosmological constraints complementary to those derived from cluster counts, and benefits from strong theoretical predictions for halo bias.

Cluster clustering, the focus of this work, has been used to constrain cosmological parameters, particularly the matter density parameter $\Omega_\mathrm{m}$ and the matter power spectrum normalisation $\sigma_8$ \citep{2018A&A...620A...1M,2023A&A...674A..80L,2024A&A...682A.148F,2024A&A...686A.196S,2025A&A...697A..93L,2026A&A...709A.102F}; to calibrate cluster masses and mass--observable relations \citep{2023A&A...674A..80L,2024A&A...686A.196S,2026arXiv261008947M}; to measure the baryon acoustic oscillation (BAO) scale \citep{2014MNRAS.442.3275V,2016MNRAS.458.1909V,2021ApJ...919..144M,2026arXiv261111587S}; and to constrain the product $f\sigma_8$ of the linear growth rate $f$ and $\sigma_8$ through template-fitting analyses with a fixed underlying matter power spectrum \citep[hereafter M21]{2021ApJ...920...13M}. The latter study provided the first estimate of $f\sigma_8$ from spectroscopic cluster clustering, conditional on a fiducial cosmological template, achieving a precision of 11\% without including BAO scales and adopting tight priors on the Alcock--Paczynski (AP) parameters.

Very recently, \citet{2026arXiv261111587S} measured the isotropic BAO signal using the \citet[hereafter WH24]{2024ApJS..272...39W} catalogue. Here, we instead use a spectroscopic \citetalias{2024ApJS..272...39W} subsample---the largest galaxy cluster sample used to date for a dedicated $f\sigma_8$ analysis---to test general relativity by jointly constraining redshift-space distortions and AP parameters from the clustering multipoles. Our analysis consists of two steps. First, we use the observed cluster number counts to constrain the relation between the observed mass proxy and the true halo mass. We then propagate the posterior distribution of the mass--proxy relation through the halo mass function and theoretical halo bias prescription to derive a prior on the effective linear bias amplitude $b_1\sigma_8$. Second, we measure the monopole and quadrupole of the redshift-space two-point correlation function (2PCF). We model these measurements using the one-loop perturbation-theory model implemented in \texttt{velocileptors}\footnote{\url{https://github.com/sfschen/velocileptors}} \citep{2020JCAP...07..062C,2021JCAP...03..100C}, incorporating the abundance-derived prior on $b_1\sigma_8$ to constrain $f\sigma_8$ and the AP parameters. Compared to \citetalias{2021ApJ...920...13M}, our analysis uses a larger sample and extends to BAO scales, allowing us to marginalise over the AP parameters using broad priors rather than imposing informative geometric constraints. 

The paper is structured as follows. In Sect. \ref{sec:methods}, we describe the data set, the estimator and random catalogue construction used to measure the 2PCF, the clustering model, and the derivation of the halo bias prior. In Sect. \ref{sec:results}, we present the measured monopole and quadrupole and the resulting cosmological constraints. Finally, in Sect. \ref{sec:conclusions}, we summarise our findings and discuss prospects for extending this analysis with next-generation facilities. We assume a flat $\Lambda$CDM fiducial cosmology with parameters set to \citet[][TT, TE, EE + lowE + lensing]{2020A&A...641A...6P}.

\section{Methods}
\label{sec:methods}

\subsection{Measurements}

\subsubsection{Data}

Our analysis uses the spectroscopic subsample of the galaxy cluster catalogue presented by \citetalias{2024ApJS..272...39W}. The full catalogue comprises approximately \num{1.58} million clusters extending to $z\sim1.5$, of which approximately 55\% are newly identified. It was constructed using optical photometry from the DESI Legacy Imaging Surveys Data Releases 9 and 10, supplemented by infrared photometry from unWISE and spectroscopic redshifts compiled from 2MASS, SDSS DR17, and the DESI Early Data Release. Clusters were identified through significant excesses in the stellar mass of galaxies within photometric-redshift intervals around massive galaxies selected as potential cluster centres. Of these clusters, \num{338841} have spectroscopic redshifts, assigned from their brightest cluster galaxies or other spectroscopically observed members. The spectroscopic subsample is therefore not uniformly selected, since cluster identification relies primarily on photometric information, while spectroscopic redshifts are available only for a subset of the identified clusters.

We select clusters with richness $\lambda_{500}\geq13$ in the redshift range $0.1 \leq z \leq 0.4$, yielding \num{85818} objects, corresponding to 25.3\% of the full spectroscopic catalogue. Following \citetalias{2024ApJS..272...39W}, $\lambda_{500}$ is defined from the total stellar mass of member galaxies within $r_{500\mathrm c}$, normalised by a characteristic stellar mass and corrected for redshift evolution. The radius $r_{500\mathrm c}$ encloses a mean density 500 times the critical density of the Universe at the cluster redshift. The adopted richness and redshift cuts are motivated by the reliability of the abundance-based halo-bias calibration and the stability of our measurements, as discussed in Sect. \ref{ssec:bias_modelling}.

\begin{figure*}
    \centering
    \includegraphics[width=0.7\textwidth]{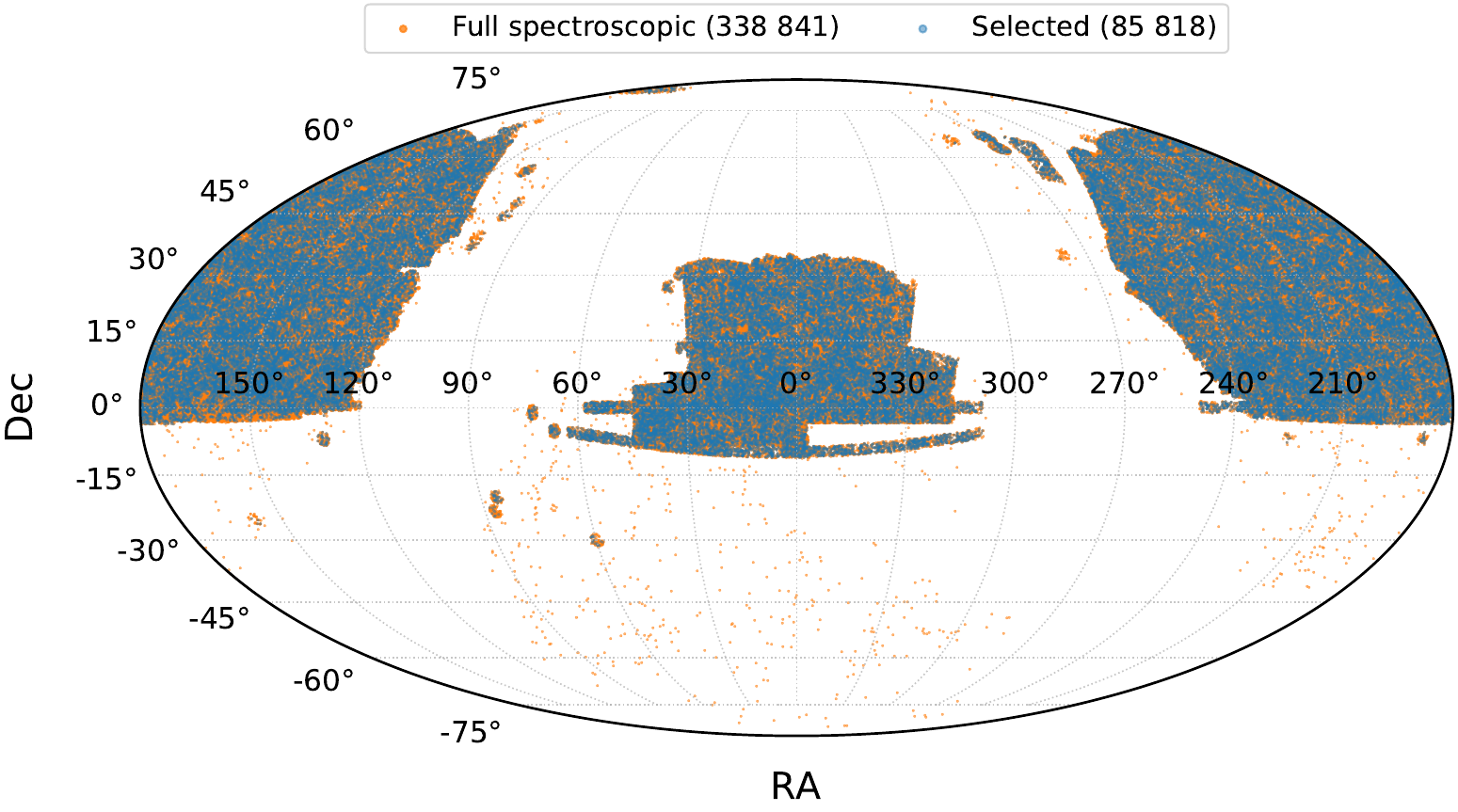}
    \caption{Angular footprint of the selected cluster sample (blue) compared to the full spectroscopic \citetalias{2024ApJS..272...39W} catalogue (orange).}
    \label{fig:fig_1a}
\end{figure*}

\begin{figure*}
    \centering
    \includegraphics[width=1\textwidth]{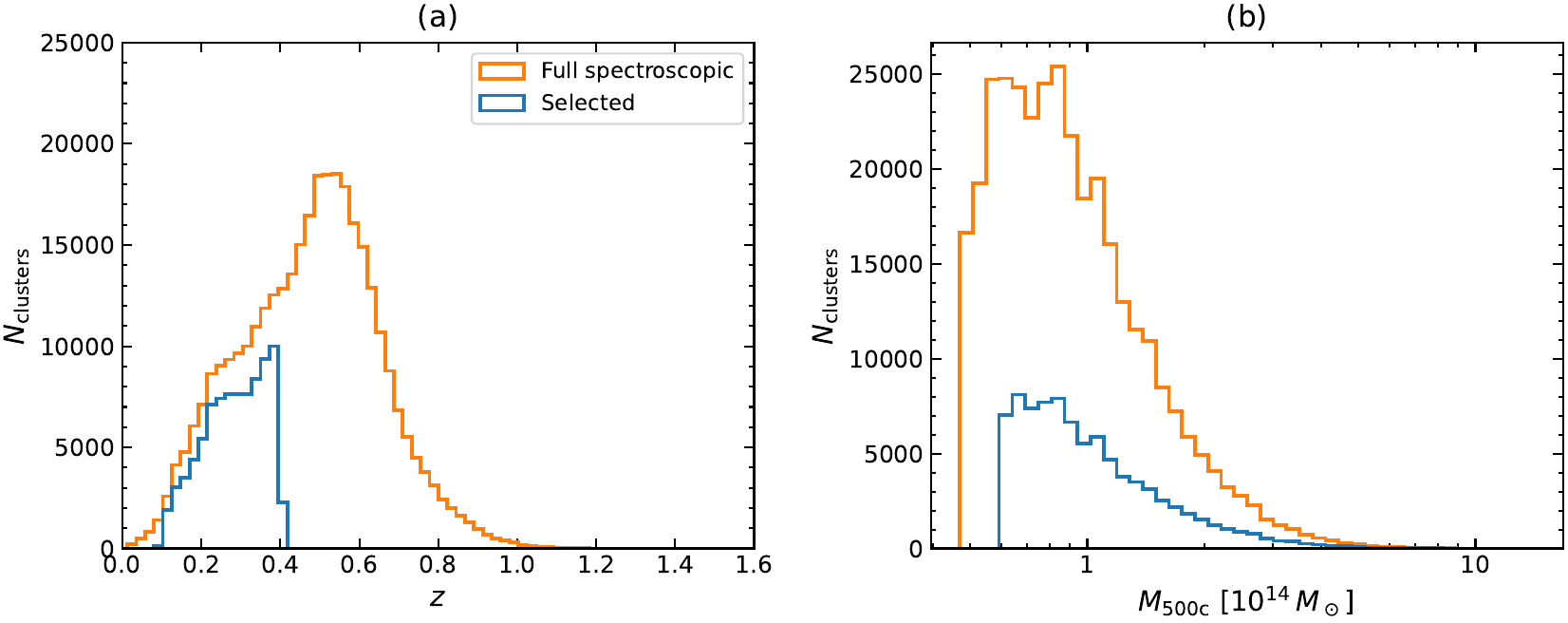}
    \caption{Redshift and mass distributions of the selected cluster sample (blue) compared to the full spectroscopic \citetalias{2024ApJS..272...39W} catalogue (orange): (a) redshift distribution and (b) $M_{500\mathrm c}$ distribution.}
    \label{fig:fig_1b}
\end{figure*}

The selected sample spans a reconstructed angular footprint of approximately \num{11300}\,deg$^2$ and has a mean effective redshift of $z_{\rm eff}\simeq0.28$. Adopting the median and pair-weighted mean as alternative definitions of the effective redshift has a negligible impact on the inferred parameters. The median catalogue-inferred mass of the selected sample is $M_{500\mathrm c}=9.2\times10^{13}\,M_\odot$, estimated from the richness--mass relation of \citetalias{2024ApJS..272...39W}, where $M_{500\mathrm c}$ denotes the mass enclosed within $r_{500\mathrm c}$. The angular distribution, redshift distribution, and catalogue-inferred mass distribution are shown in Fig. \ref{fig:fig_1a} and \ref{fig:fig_1b}. 

\subsubsection{Clustering measurements}
\label{ssec:measurement}

For the observational measurement, we divide the comoving separation range $10\leq r\leq150\,h^{-1}\mathrm{Mpc}$ into 15 logarithmically spaced bins. We measure the multipole of order $\ell$ of the redshift-space 2PCF using the direct multipole implementation of the Landy--Szalay estimator \citep{1993ApJ...412...64L},
\begin{equation}
\hat{\xi}_\ell(r)=
\frac{\mathrm{DD}^{\mathrm{norm}}_\ell(r)-2\mathrm{DR}^{\mathrm{norm}}_\ell(r)+\mathrm{RR}^{\mathrm{norm}}_\ell(r)}
{\mathrm{RR}^{\mathrm{norm}}_0(r)},
\end{equation}
where the normalised Legendre-weighted pair counts, for the unit weights adopted in our analysis, are defined as
\begin{equation}
\mathrm{XY}^{\mathrm{norm}}_\ell(r)=
\frac{2\ell+1}{N_{\mathrm{XY}}}
\sum_{p\in\mathrm{XY}(r)}
\mathcal{L}_\ell(\mu_p)\,,
\end{equation}
with $\mathrm{XY}\in\{\mathrm{DD},\mathrm{DR},\mathrm{RR}\}$ and pair-count normalisations $N_{\mathrm{DD}}=n_{\mathrm D}(n_{\mathrm D}-1)/2$, $N_{\mathrm{RR}}=n_{\mathrm R}(n_{\mathrm R}-1)/2$, and $N_{\mathrm{DR}}=n_{\mathrm D}n_{\mathrm R}$. Here, $n_{\mathrm D}$ and $n_{\mathrm R}$ denote the numbers of data and random objects, respectively, $\mathcal{L}_\ell$ is the Legendre polynomial of order $\ell$, and $\mu_p=r_{\parallel,p}/r_p$ is the cosine of the angle between the pair-separation vector and the line of sight, where $r_p$ and $r_{\parallel,p}$ denote the comoving separation of pair $p$ and its line-of-sight component, respectively. This direct multipole estimator avoids explicitly binning and integrating the two-dimensional correlation function over the line-of-sight angle. However, anisotropies in the distribution of random--random pairs can introduce angular weighting and mixing between multipoles. We have tested the estimator using simulated catalogues, finding that it accurately recovers the clustering multipoles \citep{2026A&A...713A.282E} and is numerically more stable than the integrated estimator for our application. The 2PCF measurements were obtained with \texttt{CosmoBolognaLib}\footnote{\url{https://gitlab.com/federicomarulli/CosmoBolognaLib}} \citep{2016A&C....14...35M} (version 6.1), a large set of free software C++/Python libraries.

\citetalias{2024ApJS..272...39W} provide a publicly available galaxy cluster catalogue, but not the selection function information needed to construct a random catalogue for clustering measurements. In particular, a detailed angular selection function describing the survey footprint and spatial variations in completeness is not provided, nor is a quantitative model of the spectroscopic selection function. The latter is particularly relevant because the availability of spectroscopic redshifts depends on cluster properties and sky position. We therefore reconstruct the angular selection function and estimate the radial selection directly from the observed catalogue in an approximate manner. 

Reconstructing the angular selection function requires care to distinguish variations in observational completeness from genuine cosmological clustering \citep{2026arXiv261106304F}. An overly restrictive footprint may exclude regions of genuine underdensity, whereas including regions outside the effective survey coverage may introduce spurious correlations. Moreover, a mask that follows the observed cluster distribution too closely may imprint cosmological structure onto the random catalogue, thereby suppressing the measured clustering signal. Our objective is therefore to reproduce the effective survey boundaries while minimising the influence of local fluctuations in cluster density. 

We construct the angular mask using the \citetalias{2024ApJS..272...39W} spectroscopic parent sample with $\lambda_{500}\geq10$ and restricted to $0.1\leq z \leq0.4$. To reconstruct the survey footprint, we retain all HEALPix pixels at $\mathrm{NSIDE}=512$ containing at least one parent cluster. We further restrict the mask to pixels with valid Galactic extinction information from the \citet{1998ApJ...500..525S} dust map, accessed using \texttt{dustmaps} \citep{2018JOSS....3..695M}, and stellar-density information from the Gaia EDR3 catalogue\footnote{\url{https://vizier.cds.unistra.fr/viz-bin/VizieR-3?-source=I/350/gaiaedr3}}. We detail our mask reconstruction choices and validation tests in App. \ref{ssec:mask_validation}. 

To account for potential angular variations in observational completeness, which could otherwise introduce spurious correlations into the measured 2PCF, we examine the dependence of the observed clusters in the selected sample surface density on Galactic extinction and stellar density. We calculate the number of clusters in the selected sample in each HEALPix pixel and test its correlation with each foreground template using permutation tests. For diagnostic purposes, each template is divided into 10 quantile bins, and the significance of the correlation is assessed using \num{5000} permutations, adopting a threshold of $p<0.05$.

For foreground templates showing significant correlations with cluster density, identified through a permutation test ($p<0.05$), we model their combined effect on the observed cluster counts using Poisson regression. Each selected foreground is divided into five quantile bins, and a weak $L_2$ regularisation of $10^{-4}$ is applied to stabilise the fit, excluding the intercept. The fitted model provides a relative angular density response as a function of the selected foregrounds, which we normalise to unit mean over the valid mask. We incorporate this response into the random catalogue by sampling the valid HEALPix pixels at $\mathrm{NSIDE}=512$ with probabilities proportional to their predicted relative densities. For each selected pixel, we randomly choose one of its four subpixels at $\mathrm{NSIDE}=1024$ and assign the corresponding subpixel centre as the angular position. This provides finer angular sampling while preserving the original mask and foreground-dependent selection. When neither foreground shows a significant correlation with cluster density, the valid pixels are sampled uniformly. In this way, the random catalogue reproduces the estimated foreground-dependent variations in the observed cluster density, allowing the Landy--Szalay estimator to account for these variations rather than interpreting them as cosmological clustering.

Given the heterogeneous availability of spectroscopic redshifts in the \citetalias{2024ApJS..272...39W} catalogue, we additionally investigate the dependence of spectroscopic completeness on cluster richness and potential correlations between angular position and redshift selection. We find a statistically significant richness dependence, but the residual corrections after accounting for angular selection are small, with deviations from unity of at most approximately 5\%. Tests of angular--redshift residual correlations reveal no statistically significant additional coupling. These findings motivate our adoption of a separable angular and radial selection function, without an explicit richness-dependent correction to the random catalogues.

The radial selection function is estimated from the observed redshift distribution of the richness-selected spectroscopic sample, thereby incorporating the marginal effects of richness-dependent spectroscopic completeness. We divide the selected redshift range into 60 bins and smooth the resulting histogram with a Gaussian kernel of width two bins. Random redshifts are then drawn from the smoothed distribution, uniformly within each selected bin, independently of angular position. Finally, we generate 20 random objects per observed cluster and assign unit weights to both the data and random catalogues, verifying that the measured 2PCF remains stable upon further increasing the random-to-data ratio.

\subsection{Theoretical modelling}

\subsubsection{Clustering modelling}

For our analysis, we adopt the one-loop perturbation-theory model implemented in \texttt{velocileptors} \citep{2020JCAP...07..062C,2021JCAP...03..100C}. We adopt a parametrisation that includes linear and quadratic halo bias, with higher-order Lagrangian terms set to zero. In this setting, we infer $f\sigma_8$, the linear and quadratic Eulerian bias amplitudes $b_1\sigma_8$ and $b_2\sigma_8$, an effective velocity-dispersion parameter $\sigma_{\mathrm v}$ describing additional line-of-sight damping beyond that captured by the perturbative model, and the transverse and line-of-sight AP scaling parameters $\alpha_\perp$ and $\alpha_\parallel$, respectively. To enable efficient inference of the growth rate and AP parameters, we fix $\Omega_{\mathrm m}$ and $\sigma_8$ to their fiducial values in a template-fitting approach \citepalias{2021ApJ...920...13M}. The resulting constraints on $f\sigma_8$ are therefore conditional on the assumed fiducial matter power-spectrum shape. The \texttt{velocileptors} model incorporates recent developments in perturbative redshift-space clustering while retaining a comparable number of free parameters to the extended \citet{2010PhRvD..82f3522T} model employed by \citetalias{2021ApJ...920...13M}. More specifically, the adopted model describes redshift-space distortions through an expansion in the pairwise line-of-sight velocity moments. In the plane-parallel approximation, the redshift-space power spectrum is formally expressed as
\begin{equation}
P_{\mathrm s}^{\mathrm v}(k,\mu)
=
\sum_{n=0}^{\infty}
\frac{(ik\mu)^n}{n!}\,
\Xi^{(n)}(k,\mu)\,,
\end{equation}
where $\Xi^{(n)}$ denotes the Fourier transform of the density-weighted $n$-th pairwise line-of-sight velocity moment. For a detailed derivation of the velocity-moment expansion and its perturbative contributions, we refer the reader to Sect.~2.2 of \citet{2025MNRAS.538.1789T}.

The zeroth moment describes the real-space clustering of the biased tracers, while the first and second moments incorporate the mean pairwise streaming velocity and the second pairwise velocity moment, respectively. The latter contains contributions from both velocity dispersion and mean streaming. These contributions, including their nonlinear corrections, are evaluated within one-loop perturbation theory. Thus, the coherent velocity field and its associated dispersion are already incorporated in the perturbative prediction. We parametrise the tracer bias through the Eulerian coefficients $b_1$ and $b_2$, converted to the Lagrangian coefficients entering the model using $b_1^{\mathrm L}=b_1-1$ and $b_2^{\mathrm L}=b_2-(8/21)b_1^{\mathrm L}$. We fix the tidal and cubic Lagrangian bias coefficients ($b_s=b_3=0$), effective-field-theory counterterms ($\alpha_0=\alpha_2=\alpha_4=0$), and stochastic contributions ($s_n=s_{n2}=0$) to zero. 

We incorporate AP distortions through the transformations $k'=kA(\mu)/\alpha_\perp$ and $\mu'=\mu/[FA(\mu)]$, where $F=\alpha_\parallel/\alpha_\perp$ and $A(\mu)=\sqrt{1+\mu^2(F^{-2}-1)}$. To account for residual line-of-sight damping beyond the perturbative velocity moments, we additionally adopt a Lorentzian Fingers-of-God prescription. The resulting redshift-space power spectrum is
\begin{equation}
P_{\mathrm s}(k,\mu)
=
\frac{P_{\mathrm s}^{\mathrm v}(k',\mu')}
{\alpha_\perp^2\alpha_\parallel
\left[1+(k'f\mu'\sigma_{\mathrm v})^2\right]}\,.
\end{equation}
The factor $(\alpha_\perp^2\alpha_\parallel)^{-1}$ accounts for the AP volume rescaling, while $\sigma_{\mathrm v}$ describes additional phenomenological damping beyond the total physical velocity dispersion.

Finally, the model power-spectrum multipoles are obtained as
\begin{equation}
P_\ell(k)=\frac{2\ell+1}{2}
\int_{-1}^{1}P_{\mathrm s}(k,\mu)
\mathcal{L}_\ell(\mu)\,\mathrm{d}\mu\,,
\end{equation}
where $\mathcal{L}_\ell$ denotes the Legendre polynomial of order $\ell$. We transform these multipoles into configuration space through
\begin{equation}
\xi_\ell(r)=\frac{i^\ell}{2\pi^2}
\int_0^\infty k^2P_\ell(k)j_\ell(kr)\,\mathrm{d}k\,,
\end{equation}
where $j_\ell$ is the spherical Bessel function. 

We adopt an analytic Gaussian covariance for the monopole and quadrupole of the 2PCF, assuming a constant-volume survey and Poisson shot noise. The covariance between multipoles $\ell,\ell'$ in radial bins $i,j$ is
\begin{equation}
\begin{aligned}
C_{\ell i,\ell' j}
={}&
\frac{i^{\ell+\ell'}(2\ell+1)(2\ell'+1)}
{2\pi^2 V}
\int_0^\infty \mathrm{d}k\,k^2\,
\overline{j}_{\ell,i}(k)
\overline{j}_{\ell',j}(k)\\
&\times
\int_{-1}^{1}\mathrm{d}\mu\,
\mathcal{L}_\ell(\mu)
\mathcal{L}_{\ell'}(\mu)
\left[P_{\mathrm s}^{\mathrm K}(k,\mu)^2
+\frac{2P_{\mathrm s}^{\mathrm K}(k,\mu)}{\bar n}\right]\\
&+\delta_{\ell\ell'}\delta_{ij}
\frac{2(2\ell+1)}
{V V_i\bar n^2}\,,
\end{aligned}
\label{eq:cov_model}
\end{equation}
where $V$ is the survey volume, $\bar n$ is the effective cluster number density, $V_i=4\pi(r_{i,+}^3-r_{i,-}^3)/3$ is the volume of radial shell $i$. The shell-averaged spherical Bessel functions are defined as
\begin{equation}
\overline{j}_{\ell,i}(k)
=
\frac{3}{r_{i,+}^3-r_{i,-}^3}
\int_{r_{i,-}}^{r_{i,+}}
r^2j_\ell(kr)\,\mathrm{d}r\,,
\end{equation}
where $r_{i,-}$ and $r_{i,+}$ denote the lower and upper boundaries of the $i$-th radial separation bin, respectively. The fiducial redshift-space power spectrum entering the covariance is evaluated in the linear Kaiser approximation,
\begin{equation}
P_{\mathrm s}^{\mathrm K}(k,\mu)
=
(b_1+f\mu^2)^2 P_{\mathrm m}^{\mathrm{lin}}(k,z_{\rm eff})\,,
\end{equation}
where $P_{\mathrm m}^{\mathrm{lin}}$ is the linear matter power spectrum, $k$ is the Fourier wavenumber, and $\mu$ is the cosine of the angle between the wavevector and the line of sight. We calculate this spectrum using \texttt{CAMB} through \texttt{CosmoBolognaLib}, adopting the fiducial Planck18 cosmology and evaluating it at $z_{\rm eff}=0.28$. The covariance is also obtained with \texttt{CosmoBolognaLib}.

The survey volume is calculated over $0.1\leq z \leq0.4$ using the reconstructed angular footprint of approximately $11\,300\,\mathrm{deg}^2$. For the unweighted sample, the effective number density is $\bar n=N_{\rm cl}/V$, with $N_{\rm cl}=85\,818$, corresponding to Poisson shot noise $P_{\rm shot}=1/\bar n$. The fiducial bias and growth rate entering the Kaiser spectrum are held fixed when constructing the covariance. We evaluate the Fourier integrals numerically, retaining the correlations between radial bins and between multipoles. The covariance assumes Gaussian density fluctuations and neglects survey-window mode coupling, connected non-Gaussian contributions, and super-sample covariance.

We adopt a Gaussian likelihood for the measured monopole and quadrupole,
\begin{equation}
\ln\mathcal{L}_{\rm cl} =
-\frac{1}{2}
(\hat{\boldsymbol{\xi}}-\boldsymbol{\xi})^{\rm T}
\mathbf{C}^{-1}
(\hat{\boldsymbol{\xi}}-\boldsymbol{\xi})
+\mathrm{const.}\,,
\end{equation}
where $\hat{\boldsymbol{\xi}}$ and $\boldsymbol{\xi}$ are the vectors of measured $\hat{\xi}_\ell(r)$ and modelled $\xi_\ell(r)$ multipoles, respectively, concatenated over $\ell=0,2$ and all separation bins, and $\mathbf{C}$ is the fixed Gaussian covariance in Eq. \eqref{eq:cov_model}. This likelihood choice is supported by the mock-based Gaussianity tests reported in \citet{2026A&A...713A.282E}. 
We sample the cosmological posterior using \texttt{emcee}, employing 24 walkers and \num{40000} steps for the fiducial and alternative-prior configurations, and \num{55000} steps for a restricted-quadrupole analysis which will be discussed in Sect. \ref{ssec:quadrupole}. We discard the first \num{5000} steps as burn-in.

\subsubsection{Clustering bias prediction}
\label{ssec:bias_modelling}

We predict the effective linear bias of the \citetalias{2024ApJS..272...39W} cluster sample using the halo mass function, a calibrated relation between observed and true halo masses, and the theoretical halo bias. We define the observed mass proxy $M_{\rm p}$ using the richness--mass calibration of \citetalias{2024ApJS..272...39W},
\begin{equation}
\log_{10}
\left(\frac{M_{\rm p}}{10^{14}M_\odot}\right)
=
0.97\log_{10}\lambda_{500}-1.3\,.
\end{equation}
This relation provides an estimate of $M_{500\mathrm{c}}$ from the observed cluster richness, with masses expressed in $M_\odot$. For the abundance and halo-bias calculations, we convert the mass proxy to units of $h^{-1}M_\odot$ by multiplying its numerical value by $h=0.674$. 

The quantity of interest is the effective bias amplitude,
\begin{equation}
    (b_1\sigma_8)_{\mathrm{eff}}
    = \frac{1}{N_{\mathrm{cl}}}
    \sum_{i=1}^{N_{\mathrm{cl}}}
    \langle b_1\rangle(M_{{\rm p},i},z_i)
    \sigma_8(z_i)\,,
\end{equation}
where $N_{\mathrm{cl}}$ is the number of observed clusters, $M_{{\rm p},i}$ and $z_i$ are their mass proxies and redshifts, respectively, and $\sigma_8(z)$ is the linear matter fluctuation amplitude at redshift $z$. The quantity $\langle b_1\rangle(M_{\rm p},z)$ is the expected linear halo bias conditional on the observed mass proxy and redshift, given by
\begin{equation}
    \langle b_1\rangle(M_{\rm p},z)
    = \int b_1(M,z)\,P(M\mid M_{\rm p},z)\,\mathrm{d}M\,,
\end{equation}
where $M$ denotes the true halo mass and $b_1(M,z)$ is the theoretical linear halo bias, evaluated using the \citet[hereafter T10]{2010ApJ...724..878T} prescription. This prescription expresses the linear halo bias as a function of peak height with fitting coefficients calibrated against N-body simulations that depend on the spherical-overdensity halo-mass definition. The authors report approximately $6\%$ scatter in their best fit relation. This represents a theoretical calibration uncertainty that is not captured by the statistical uncertainty of our abundance-based prediction. To assess the sensitivity of our cosmological constraints on this uncertainty, we additionally perform the clustering inference with a uniform bias prior. 

The conditional true-mass distribution is obtained through Bayes' theorem
\begin{equation}
    P(M\mid M_{\rm p},z)
    =
    \frac{
    P(M_{\rm p}\mid M,z)\,n(M,z)
    }{
    \int P(M_{\rm p}\mid M',z)\,n(M',z)\,\mathrm{d}M'
    }\,,
\end{equation}
where $n(M,z)=\mathrm{d}n/\mathrm{d}M$ is the differential halo mass function, evaluated using the \citet{2008ApJ...688..709T} prescription. The mass function acts as the prior distribution of true halo masses, accounting for the fact that, owing to the steeply declining abundance of massive haloes, more low-mass objects scatter towards high observed masses than vice versa. 

To determine the likelihood $P(M_{\rm p}\mid M,z)$, we model the logarithm of the observed mass proxy as a Gaussian random variable at fixed true mass and redshift following \citet{2018ApJ...854..120M}. 
Defining $x=\log_{10}[M_{\rm p}/(h^{-1}M_\odot)]$ and $m=\log_{10}[M/(h^{-1}M_\odot)]$, we adopt
\begin{equation}
    P(x\mid m,z)
    =
    \frac{1}{\sqrt{2\pi}\,\sigma(m,z)}
    \exp\left[
    -\frac{(x-\mu(m,z))^2}{2\sigma^2(m,z)}
    \right]\,.
\end{equation}
The mean relation is parametrised as $\mu(m,z)=A+\alpha m+\gamma(z-z_{\rm piv})$, while the logarithmic scatter is $\sigma(m,z)=B+\beta m+\eta(z-z_{\rm piv})$. We adopt $z_{\rm piv}=0.2$ as a reference redshift within the sample range. The six parameters $(A,\alpha,B,\beta,\gamma,\eta)$ describe the normalisation, mass dependence, and redshift evolution of the mean relation and its intrinsic scatter. The scatter is required to remain positive throughout the fitted mass and redshift range.

We estimate the spectroscopic selection function directly from the \citetalias{2024ApJS..272...39W} parent catalogue, using both spectroscopic and photometric-only clusters. The latter provide the reference population needed to determine the fraction of parent clusters with available spectroscopic redshifts as a function of richness, redshift, and sky position. We model this probability as
\begin{equation}
    S_{\rm spec}(\lambda,z,\mathrm{RA},\mathrm{Dec})
    =
    \frac{1}{1+\exp[-g(\lambda,z,\mathrm{RA},\mathrm{Dec})]}\,,
\end{equation}
where $\mathrm{RA}$ and $\mathrm{Dec}$ denote right ascension and declination. The function $g$ is represented by a tensor product of cubic B-splines in $\log_{10}\lambda$ and $z$, supplemented by 14 angular basis functions describing spatial variations in spectroscopic completeness. The angular parametrisation is motivated by out-of-sample validation: increasing the number of angular basis functions from 6 to 14 improves the held-out Bernoulli log-loss, while tests of additional angular- richness and angular--redshift interactions show no improvement in predictive performance. We determine the coefficients by minimising the Bernoulli negative log-likelihood, with a second-difference regularisation penalty of strength $10^{-5}$ applied along the richness and redshift directions. We validate the model using a held-out subset comprising 20\% of the parent catalogue. As shown in Fig. \ref{fig:fig_3}a--b, the predicted spectroscopic completeness agrees with the observed fractions as functions of richness and redshift in the held-out sample. Following validation, the selection model is refitted using the full parent sample.

For the abundance calculation, we average the fitted spectroscopic selection function over the survey footprint
\begin{equation}
    \overline{S}_{\rm spec}(\lambda,z)
    =
    \frac{1}{\Omega_{\rm surv}}
    \int_{\Omega_{\rm surv}}
    S_{\rm spec}(\lambda,z,\mathrm{RA},\mathrm{Dec})\,d\Omega\,,
    \label{eq:S_spec}
\end{equation}
where $\Omega_{\rm surv}$ denotes the solid angle of the reconstructed footprint. We evaluate this average numerically using \num{20000} angular positions sampled uniformly within the footprint.

We constrain the mass--proxy relation by fitting the observed abundance of spectroscopic clusters as a function of redshift and proxy mass. We evaluate the abundance in four redshift bins over $0.1\leq z\leq0.4$ and 17 logarithmically spaced proxy-mass bins. For a redshift interval $[z_a,z_b]$ and an observed logarithmic mass interval $[x_a,x_b]$, the expected number of spectroscopic clusters is
\begin{eqnarray}
    N_{\rm pred}
    &=& \int_{z_a}^{z_b} \mathrm{d}z\,\frac{\mathrm{d}V}{\mathrm{d}z}
    \int \mathrm{d}\ln M\,\frac{\mathrm{d}n}{\mathrm{d}\ln M}(M,z)
    \nonumber\\
    && \times \int_{x_a}^{x_b} \mathrm{d}x\,
    P(x\mid m(M),z)\,
    \overline{S}_{\rm spec}(\lambda(x),z)\,.
\end{eqnarray}
where $\mathrm{d}V/\mathrm{d}z$ is the comoving volume element over the reconstructed survey footprint and $\overline{S}_{\rm spec}(\lambda,z)$ is the sky-averaged probability that a cluster of richness $\lambda$ and redshift $z$ has a spectroscopic redshift in Eq. \eqref{eq:S_spec}.

In our abundance calibration, we approximate the detection completeness of the WH24 photometric parent catalogue as unity within the adopted selection cuts. This assumption is motivated by the catalogue validation presented by \citetalias{2024ApJS..272...39W}, who report recovery fractions of approximately 95\% relative to redMaPPer, 94\% relative to \citet{2022MNRAS.513.3946W} at $z<1$, and 90--92\% relative to several independent X-ray-selected cluster samples. Importantly, the authors report good recovery even at masses as low as $M_\mathrm{500c}\sim5\times10^{13}\,M_\odot$, below the approximate mass threshold of our sample, $M_\mathrm{500c}\gtrsim6\times10^{13}\,M_\odot$. Together with our restriction to $0.1\leq z \leq0.4$, these findings motivate treating the parent catalogue as approximately complete. To assess the sensitivity of our results to this assumption, we consider alternative parent detection completeness functions taking values of 80\%, 90\%, and 95\% at $\lambda_{500}=13$ and increasing linearly to unity at $\lambda_{500}=50$. For each scenario, we recalibrate the abundance-derived halo-bias prior and repeat the clustering inference. We find that the inferred $f\sigma_8$ remains consistent with the fiducial result within its statistical uncertainty in all scenarios.

For computational efficiency, the halo mass function, volume element, and spectroscopic selection function are integrated over redshift within each bin, while the mass--proxy relation is evaluated at the bin centre. Within each proxy-mass bin, the selection function is evaluated at the corresponding richness derived from the \citetalias{2024ApJS..272...39W} calibration. The probability of a true halo mass contributing to each observed mass bin is computed analytically from the Gaussian cumulative distribution function.

We fit the mass--proxy relation using a Gaussian likelihood for the observed cluster counts,
\begin{equation}
\begin{aligned}
\ln\mathcal{L} = -\frac12 \Big[
&(\mathbf N^{\rm obs}-\mathbf N^{\rm pred})^{\rm T}
\mathbf C_\mathrm{N}^{-1}
(\mathbf N^{\rm obs}-\mathbf N^{\rm pred})\\
&+\ln|\mathbf C_\mathrm{N}|
\Big]+\mathrm{const.}\,,
\end{aligned}
\end{equation}
where $\mathbf{N}^{\rm obs}$ and $\mathbf{N}^{\rm pred}$
are the observed and predicted cluster counts across
the redshift and proxy-mass bins, respectively. The covariance matrix includes both Poisson shot noise and sample variance
\begin{equation}
    (C_\mathrm{N})_{jk,j'k'}
    =
    N_{jk}^{\rm pred}\delta_{jj'}\delta_{kk'}
    +
    Q_{jk}Q_{j'k'}S_{jj'}\,,
\end{equation}
where $j$ and $k$ label the redshift and proxy-mass bins, respectively, and $Q_{jk}=(Nb)_{jk}^{\rm pred}+f(z_j)N_{jk}^{\rm pred}$. Here, $(Nb)_{jk}^{\rm pred}$ denotes the predicted counts weighted by linear halo bias, and $S_{jj'}$ is the covariance of matter density fluctuations between redshift bins. We compute $S_{jj'}$ using the reconstructed survey angular window, incorporating linear redshift-space distortions through the effective-power-spectrum approximation of \citet{2026A&A...709A.102F}.

We compare three nested parametrisations of the mass--proxy relation: a four-parameter model without redshift evolution ($\gamma=\eta=0$), a five-parameter model allowing evolution of the mean relation ($\eta=0$), and a six-parameter model allowing both the mean and scatter to evolve with redshift. Their maximum-likelihood solutions are obtained through multistart optimisation and compared using the Akaike and Bayesian information criteria. Both criteria strongly favour redshift evolution of the mean relation. Conversely, allowing for additional redshift evolution of the intrinsic scatter is not statistically justified. We therefore adopt the five-parameter model for our fiducial analysis. 

For each set of mass--proxy parameters, the corresponding Gaussian distribution $P(x\mid m,z)$ determines the probability of a halo contributing to each observed proxy-mass bin and hence the predicted cluster counts. Both the predicted counts and their covariance depend on these parameters and are updated at each likelihood evaluation. Therefore, the observed abundance jointly constrains the normalisation, mass dependence, intrinsic scatter, and redshift evolution of the mass--proxy relation. We adopt uniform priors on the five mass--proxy parameters, with $A+14\alpha-14\sim\mathcal{U}(-1,1)$, $\alpha\sim\mathcal{U}(0.1,1.5)$, $B+14\beta\sim\mathcal{U}(0.02,0.8)$, $\beta\sim\mathcal{U}(-0.5,0.5)$, and $\gamma\sim\mathcal{U}(-2,2)$. We additionally require the intrinsic scatter $\sigma(m,z)=B+\beta m$ to exceed $0.02$ throughout the fitted true-mass range, while fixing $\eta=0$. The inferred $b_1\sigma_8$ is robust to these choices. We sample the five-parameter posterior using \texttt{emcee}, employing 24 walkers and \num{5000} steps, discarding the first \num{1000} steps as burn-in.

Finally, we propagate the full posterior distribution of the mass--proxy parameters into the effective linear bias prediction. For each posterior draw, we recompute $P(M\mid M_\mathrm{p},z)$ using the corresponding mass--proxy relation and the \citet{2008ApJ...688..709T} halo mass function. We then evaluate the conditional mean halo bias using the \citetalias{2010ApJ...724..878T} prescription and calculate $(b_1\sigma_8)_\mathrm{eff}$ from the observed joint distribution of proxy masses and redshifts, using 20 redshift bins and 30 logarithmic mass bins. We adopt $\sigma_8(z)=D(z)\sigma_{8,0}$, where $D(z)$ is the linear growth factor normalised to unity at $z=0$, adopting the fiducial \citet{2020A&A...641A...6P} cosmology. Repeating this calculation for all posterior draws yields the abundance-derived posterior distribution of $(b_1\sigma_8)_\mathrm{eff}$, incorporating the uncertainties and correlations among the fitted mass--proxy parameters. We approximate this distribution by a Gaussian prior on $b_1\sigma_8$ in the subsequent clustering inference, while $b_2\sigma_8$ is independently constrained by the clustering measurement. We validate the reliability of the bias prediction in App. \ref{ssec:bias_validation}. 

We determine the richness and redshift limits of the selected cluster sample by examining the agreement between the observed and predicted cluster abundances and the reliability of the spectroscopic selection function for different sample selections. We adopt a lower redshift limit of $z_{\min}=0.1$, as extending the sample to lower redshifts results in a substantial overprediction of the observed cluster abundance. We then explore different combinations of the minimum richness $\lambda_{\min}$ and maximum redshift $z_{\max}$. We find that lowering the richness threshold below $\lambda_{\min}=13$ introduces systematic discrepancies between the observed and predicted proxy-mass distributions, particularly at the low-mass end. Conversely, extending the sample beyond $z_{\max}=0.4$ leads to a deterioration in the agreement between the observed and predicted spectroscopic completeness. We therefore adopt $\lambda_{\min}=13$ and $0.1\leq z \leq0.4$, which provide a suitable compromise between sample size and modelling reliability, yielding the final sample of \num{85818} clusters.

\section{Results}
\label{sec:results}

\begin{table}
\centering
\caption{Prior distributions and marginalised posterior constraints.}
\label{tab:priors_posteriors}
\renewcommand{\arraystretch}{1.5}
\begin{tabular}{lcc}
\hline\hline
Parameter & Prior & Posterior (68\% CI) \\
\hline
$f\sigma_8$ & $\mathcal{U}(0,1)$ & $0.45\pm0.08$ \\
$b_1\sigma_8$ & $\mathcal{N}(1.34,0.01^2)$ & $1.34\pm0.01$ \\
$b_2\sigma_8$ & $\mathcal{U}(-10,10)$ & $-1.01^{+0.28}_{-0.26}$ \\
$\sigma_\mathrm{v}\;[h^{-1}\mathrm{Mpc}]$ & $\mathcal{U}(0,100)$ & $0.96^{+1.24}_{-0.67}$ \\
$\alpha_\perp$ & $\mathcal{U}(0.5,1.5)$ & $0.96\pm0.04$ \\
$\alpha_\parallel$ & $\mathcal{U}(0.5,1.5)$ & $0.99\pm0.04$ \\
\hline
\end{tabular}
\end{table}

\begin{figure}
    \centering
    \includegraphics[width=0.5\textwidth]{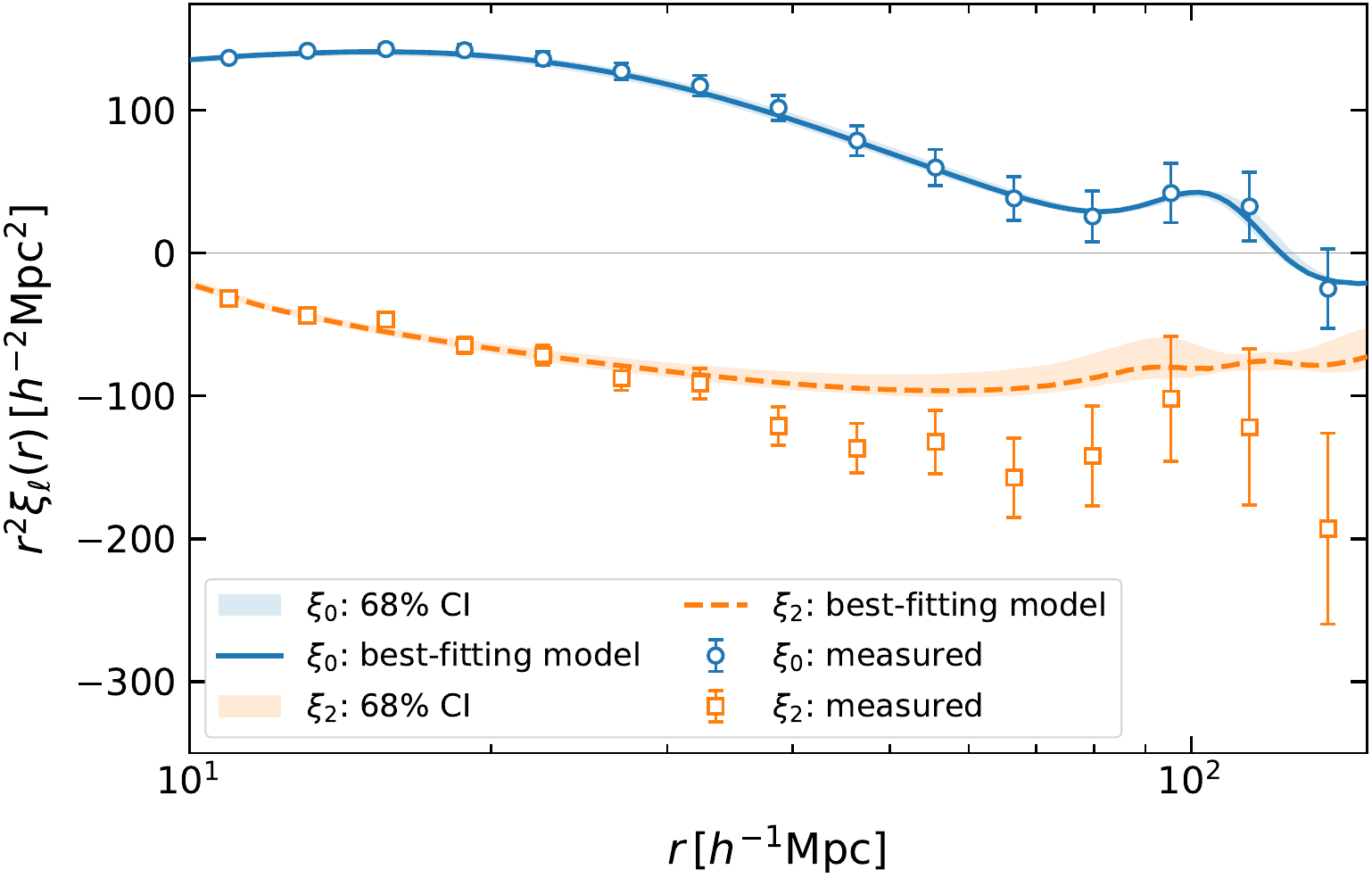}
    \caption{
    Redshift-space monopole ($\ell=0$, circles) and quadrupole ($\ell=2$, squares) of the two-point correlation function measured from our spectroscopic cluster sample with $\lambda_{500}\geq13$ and $0.1\leq z \leq0.4$. The solid and dashed curves show the respective best-fitting clustering-model predictions over $10\leq r\leq150\,h^{-1}\mathrm{Mpc}$, with shaded regions indicating the pointwise 68\% posterior credible intervals (CI). Error bars correspond to the diagonal elements of the Gaussian covariance matrix, while the fit accounts for the full covariance, including correlations between radial bins and multipoles. The joint fit yields $\chi^2_{\mathrm{red}}\simeq0.94$.
    }
    \label{fig:fig_4}
\end{figure}

\begin{figure}
    \centering
    \includegraphics[width=0.5\textwidth]{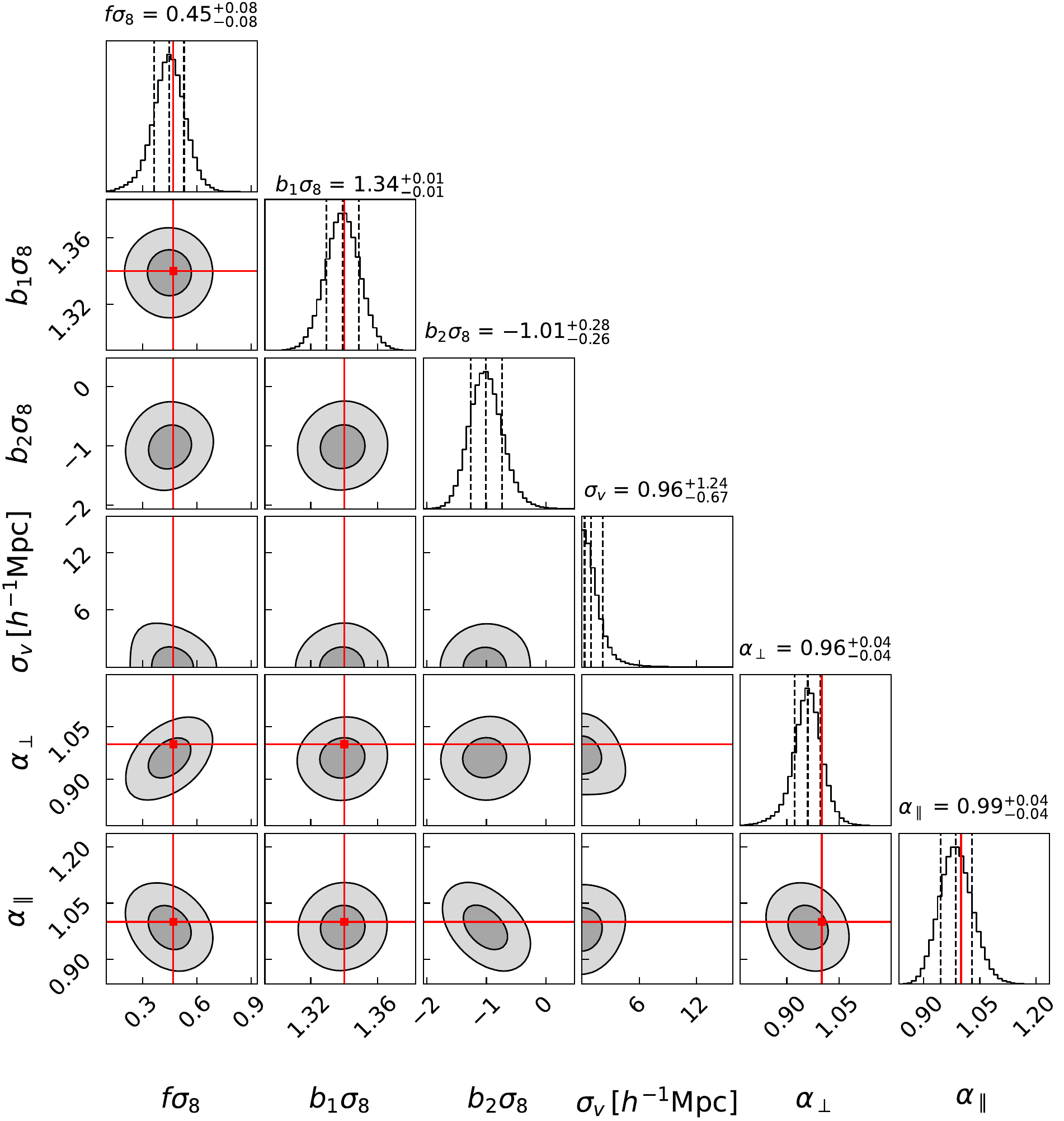}
    \caption{Marginalised posterior distributions of $f\sigma_8$, the Eulerian $b_1\sigma_8$ and $b_2\sigma_8$, $\sigma_\mathrm{v}$, $\alpha_\perp$, and $\alpha_\parallel$ inferred from the \citetalias{2024ApJS..272...39W} redshift-space monopole and quadrupole over $10\leq r\leq150\,h^{-1}\mathrm{Mpc}$, for $\lambda_{500}\geq13$ and $0.1\leq z \leq0.4$. The analysis adopts a Gaussian prior of $b_1\sigma_8=1.34\pm0.01$ derived from the cluster abundance, while the AP parameters are assigned broad priors. Contours enclose the 68\% and 95\% credible regions, and the quoted one-dimensional constraints correspond to the median and 68\% credible intervals. The red lines indicate the reference values for $f\sigma_8$, $b_1\sigma_8$, $\alpha_\perp$, and $\alpha_\parallel$.}
    \label{fig:fig_5}
\end{figure}

Fig. \ref{fig:fig_4} presents the measured monopole and quadrupole of the redshift-space 2PCF, together with the best-fitting clustering model. The model provides an adequate joint fit over $10\leq r\leq150\,h^{-1}\mathrm{Mpc}$, with $\chi^2_{\rm red}\simeq0.94$. By jointly modelling these multipoles, we constrain the growth rate $f\sigma_8$ and the AP parameters $\alpha_\perp$ and $\alpha_\parallel$, allowing for both dynamical and geometric contributions to the observed redshift-space anisotropies. We obtain $f\sigma_8=0.45\pm0.08$ at an effective redshift of $z_{\rm eff}\simeq0.28$, with $\alpha_\perp=0.96\pm0.04$ and $\alpha_\parallel=0.99\pm0.04$. These constraints are consistent with the predictions of general relativity within the adopted fiducial cosmology. We further infer $b_1\sigma_8=1.34\pm0.01$, $b_2\sigma_8=-1.01^{+0.28}_{-0.26}$, and $\sigma_{\mathrm v}=0.96^{+1.24}_{-0.67}\,h^{-1}\mathrm{Mpc}$. The marginalised posterior distributions are shown in Fig. \ref{fig:fig_5}, while Table \ref{tab:priors_posteriors} summarises the adopted priors and corresponding posterior constraints for all six inferred parameters.

\subsection{Large-scale quadrupole}\label{ssec:quadrupole}

The quadrupole measurement exhibits a coherent negative offset relative to the best-fitting model of up to approximately $2\sigma$ per separation bin at $r>33\,h^{-1}\mathrm{Mpc}$. Similar behaviour has been reported in previous cluster-clustering \citepalias{2021ApJ...920...13M} and galaxy-clustering analyses \citep{2016MNRAS.456.4156K,2025A&A...698A.170P}, where correlations between separation bins have been discussed as a possible explanation for coherent deviations. The feature persists across different angular-mask resolutions, while the selection-function validation tests described in Sect. \ref{ssec:measurement} show no significant residual angular--redshift coupling or preferential imprinting of cluster overdensities onto the random catalogue.

To quantify the statistical significance of this residual, we consider the eight quadrupole measurements at $r>33\,h^{-1}\mathrm{Mpc}$, corresponding to separations between approximately $40$ and $140\,h^{-1}\mathrm{Mpc}$. Accounting for their covariance and correlations with the remaining monopole and quadrupole measurements, we find a coherent negative deviation of $2.7\sigma$ from the fiducial model. We assess its significance using 100 correlated Gaussian mock measurements generated from the best-fitting full-range model, refitting the model parameters to each realisation. For the fixed threshold of $33\,h^{-1}\mathrm{Mpc}$, the observed offset has an empirical two-sided probability of $p=0.02$. However, the threshold of $33\,h^{-1}\mathrm{Mpc}$ was chosen after inspecting the measured quadrupole, potentially enhancing the apparent significance of the residual. To account for this effect, we consider all 12 possible starting bins for a large-scale quadrupole residual, spanning approximately $10$ to $80\,h^{-1}\mathrm{Mpc}$, while requiring at least four consecutive measurements extending to the largest fitted separation. For both the observations and each mock realisation, we calculate the coherent offset for all 12 scale selections and retain the largest absolute deviation when evaluating the two-sided significance. The most negative observed deviation is obtained for the eight bins starting at approximately $38.7\,h^{-1}\mathrm{Mpc}$. Using 5000 Gaussian mock realisations, we estimate a two-sided probability of $p\simeq0.04$ after accounting for the selection of the starting scale. Thus, although the large-scale quadrupole residual exhibits a modest statistical significance, it does not by itself establish the presence of an observational systematic or modelling deficiency. Given the consistency of the cosmological constraints under the quadrupole scale cut, we therefore retain these measurements in the fiducial analysis while acknowledging the unresolved origin of the residual.

We additionally assess the sensitivity of the cosmological constraints to the large-scale quadrupole by repeating the \texttt{velocileptors} analysis with quadrupole measurements restricted to $r\leq33\,h^{-1}\mathrm{Mpc}$, retaining the full monopole fitting range and the corresponding Gaussian covariance submatrix. We obtain $f\sigma_8=0.26^{+0.13}_{-0.14}$, $b_1\sigma_8=1.34\pm0.01$, $b_2\sigma_8=-1.11^{+0.34}_{-0.26}$, $\sigma_{\mathrm v}=3.90^{+8.61}_{-2.90}\;h^{-1}\mathrm{Mpc}$, $\alpha_\perp=0.88^{+0.06}_{-0.07}$, and $\alpha_\parallel=1.04^{+0.07}_{-0.06}$. These constraints remain compatible with the fiducial results, although their uncertainties increase following the removal of quadrupole information. The marginalised $f\sigma_8$ posterior shifts towards lower values, accompanied by stronger degeneracies with $\sigma_{\mathrm v}$ and $\alpha_\parallel$, potentially reflecting prior-volume effects in less constraining analyses \citep{2026A&A...713A.282E}. The pronounced high-$\sigma_{\mathrm v}$ tail also exhibits slow MCMC mixing, introducing additional uncertainty in the restricted-fit posterior intervals. To determine whether the parameter shifts are statistically unusual, we independently optimise the posterior for the full and restricted fitting ranges in each of the 100 Gaussian mocks. The observed maximum-a-posteriori (MAP) shift, $\Delta(f\sigma_8)=0.099$, is consistent with the mock distribution, which has a mean of $0.068$ and a standard deviation of $0.170$, yielding an empirical two-sided probability of $p=0.92$. Similarly, the observed AP parameter shifts, $\Delta\alpha_\perp=0.041$ and $\Delta\alpha_\parallel=-0.023$, yield $p=0.91$ and $p=0.94$, respectively. Thus, the changes in the cosmological parameters following the quadrupole scale cut are compatible with the statistical fluctuations expected under the adopted model.

Finally, we investigate whether survey-window-induced multipole mixing or integral-constraint effects could explain the quadrupole offset. Applying a mixing matrix derived from the angular distribution of random--random pairs produces negligible corrections on the affected scales, while fitting an integral-constraint-like constant offset, including its coupling to the quadrupole, yields no significant improvement in the fit. Neither effect accounts for the observed residual. Given the acceptable overall goodness of fit, the modest statistical significance of the coherent quadrupole offset after accounting for scale selection, and the consistency of the cosmological constraints under the scale cut, we retain the full separation range in our fiducial analysis while acknowledging that the origin of the large-scale quadrupole residual remains unresolved.

\subsection{Quadratic halo bias}

Our results favour a significantly non-zero Eulerian quadratic bias, with $b_2\sigma_8=-1.01^{+0.28}_{-0.26}$, compared to $b_2\sigma_8\simeq-0.16$ predicted by the empirical $b_2(b_1)$ relation of \citet[][Eq.~5.2]{2016JCAP...02..018L}, evaluated at the effective linear bias of our cluster sample. To assess the sensitivity of our cosmological constraints to this discrepancy, we repeat the inference imposing a tight Gaussian prior, $b_2\sigma_8\sim\mathcal{N}(-0.16,0.01^2)$, effectively fixing the quadratic bias to its theoretical expectation. We obtain $f\sigma_8=0.37^{+0.10}_{-0.12}$, statistically consistent with our fiducial constraint of $0.45\pm0.08$. The goodness of fit deteriorates from $\chi^2_{\mathrm{red}}\simeq0.94$ to $\chi^2_{\mathrm{red}}\simeq1.29$. The inferred velocity-dispersion parameter increases to $\sigma_{\mathrm v}=3.29^{+2.87}_{-1.45}\;h^{-1}\mathrm{Mpc}$, while the AP parameters shift towards lower values, $\alpha_\perp=0.91^{+0.06}_{-0.07}$ and $\alpha_\parallel=0.95^{+0.06}_{-0.05}$. These results indicate that the inferred growth rate is relatively stable under this restrictive quadratic-bias assumption, whereas the velocity-dispersion and geometric parameters exhibit greater sensitivity, suggesting degeneracies within the adopted clustering model.

In our adopted \texttt{velocileptors} parametrisation, higher-order Lagrangian terms are fixed to zero. We have explored allowing these additional parameters to vary, but find that the data provide insufficient constraining power to determine their amplitudes independently. Consequently, the inferred $b_2$ may effectively absorb contributions associated with these additional bias terms, rather than corresponding directly to the quadratic bias calibrated in simulations. Differences in halo-mass definitions and the mass and redshift distribution of the selected cluster sample may also contribute to the discrepancy. The inferred quadratic bias remains consistent between our fiducial and restricted-scale analyses, indicating that its departure from the \citet{2016JCAP...02..018L} relation is not driven by the large-scale quadrupole residuals.

\subsection{Lorentzian damping term}

The posterior distribution of the velocity-dispersion parameter $\sigma_{\mathrm v}$ favours relatively low values, despite the finite redshift uncertainties of the spectroscopic cluster sample. This behaviour does not imply the absence of peculiar velocities, as the contribution of the large-scale velocity field, including its linear one-dimensional velocity dispersion, is already incorporated in the perturbative predictions of \texttt{velocileptors}. The additional parameter $\sigma_{\mathrm v}$ describes phenomenological damping beyond that captured by the perturbative model, including unresolved small-scale velocity contributions and potentially redshift uncertainties. Its preference for low values therefore indicates that the fiducial measurements do not require substantial additional damping within the adopted parametrisation. However, imposing a tight theoretical prior on $b_2\sigma_8$ shifts the posterior towards higher $\sigma_{\mathrm v}$, demonstrating that the inferred residual damping depends on the adopted quadratic-bias prescription. We therefore do not interpret $\sigma_{\mathrm v}$ as an independent measurement of the physical velocity dispersion.

\subsection{Alcock-Paczynski parameters}

The AP parameters are also of particular interest, as geometric distortions can partially mimic the anisotropies induced by redshift-space distortions, introducing degeneracies with the growth rate. Our fiducial constraints, $\alpha_\perp=0.96\pm0.04$ and $\alpha_\parallel=0.99\pm0.04$, are consistent with unity. These constraints are consistent with both the isotropic BAO measurement obtained from a different spectroscopic selection of the WH24 catalogue by \citet{2026arXiv261111587S} and the DESI DR2 Bright Galaxy Survey BAO measurement at a comparable effective redshift \citep{2025PhRvD.112h3515A}. The quadrupole scale-cut analysis and the associated Gaussian-mock tests further indicate that including the large-scale quadrupole improves the precision of the geometric constraints without producing parameter shifts that are statistically unusual under the adopted model and covariance.

To assess the impact of the AP prior assumptions and facilitate comparison with \citetalias{2021ApJ...920...13M}, we repeat our analysis adopting Gaussian priors centred on unity with a standard deviation of $0.01$ for both $\alpha_\perp$ and $\alpha_\parallel$, as in their work where the BAO peak could not be measured robustly. Over our fiducial separation range of $10\leq r\leq150\,h^{-1}\mathrm{Mpc}$, we obtain $f\sigma_8=0.52\pm0.03$, compared to $f\sigma_8=0.45\pm0.08$ with broad AP priors. The tighter priors reduce the growth-rate uncertainty by a factor of approximately $2.5$, illustrating the impact of geometric degeneracies on the inferred growth rate. Restricting the fitting range to $10\leq r\leq80\,h^{-1}\mathrm{Mpc}$, as in \citetalias{2021ApJ...920...13M}, yields $f\sigma_8=0.52\pm0.03$. This is consistent with their estimate of $f\sigma_8=0.44\pm0.05$ at $z_{\mathrm{eff}}=0.275$. The agreement is obtained despite differences in the cluster samples, clustering statistics, and theoretical models. These comparisons demonstrate that the growth-rate constraint is stable with respect to the maximum fitting scale under the same AP priors, while the assumed geometric priors substantially affect its precision and central value.

\subsection{Sensitivity to linear halo bias prior}

We investigate the contribution of the abundance-derived linear bias prior to the cosmological constraints by replacing the Gaussian prior on $b_1\sigma_8$ with a uniform prior, while keeping the remaining analysis settings unchanged. This test also provides an assessment of the sensitivity of our results to uncertainties in the theoretical halo-bias calibration. The approximately 6\% scatter reported by \citetalias{2010ApJ...724..878T} corresponds to an uncertainty of approximately 0.08 in our effective $b_1\sigma_8$ prediction. We obtain $f\sigma_8=0.42^{+0.09}_{-0.10}$, $b_1\sigma_8=1.31^{+0.05}_{-0.07}$, $b_2\sigma_8=-1.13^{+0.34}_{-0.35}$, $\sigma_{\mathrm v}=0.95^{+1.50}_{-0.67}\;h^{-1}\mathrm{Mpc}$, $\alpha_\perp=0.93^{+0.05}_{-0.07}$, and $\alpha_\parallel=0.99\pm0.04$. The linear bias inferred from clustering alone agrees with the abundance-based prediction of $b_1\sigma_8=1.34\pm0.01$, providing an independent consistency check between the two approaches. Incorporating the abundance-derived prior reduces the average uncertainty on $f\sigma_8$ from $0.10$ to $0.08$, corresponding to a 20\% improvement in precision, while leaving the inferred growth rate statistically consistent. This behaviour is consistent with \citetalias{2021ApJ...920...13M}, who similarly found that replacing an informative linear-bias prior with a uniform prior increased the uncertainty on $f\sigma_8$ from $0.05$ to $0.07$, while leaving the inferred growth rate statistically consistent. 

\begin{figure*}
    \centering
    \includegraphics[width=\textwidth]{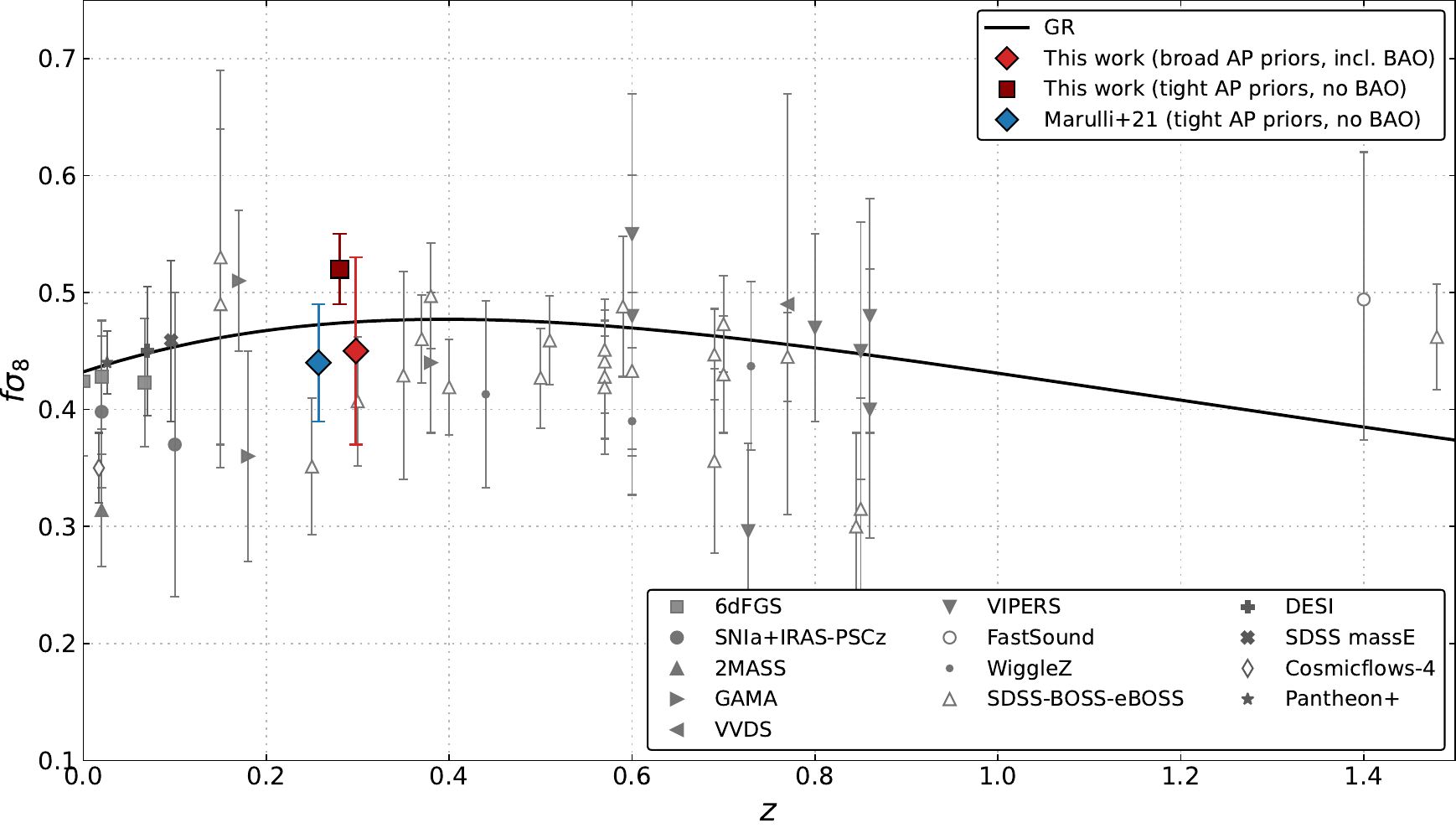}
    \caption{Compilation of representative estimates of $f\sigma_8$ as a function of redshift. The black solid curve represents the Planck18 $\Lambda$CDM prediction assuming general relativity. The galaxy cluster clustering constraints from this work and \citetalias{2021ApJ...920...13M} are highlighted separately. Our fiducial analysis (red diamond) includes BAO scales, extending to $150\,h^{-1}\mathrm{Mpc}$, and adopts broad uniform priors on the AP parameters. We additionally show the constraint obtained by restricting the fitting range to $10$--$80\,h^{-1}\mathrm{Mpc}$ and imposing Gaussian AP priors centred on unity with a standard deviation of $0.01$ (dark-red square), matching the scale range and geometric prior assumptions of \citetalias{2021ApJ...920...13M} (blue diamond). The comparison illustrates the sensitivity of growth-rate constraints to the adopted geometric priors. The three cluster-clustering estimates are slightly offset horizontally for visual clarity. The reference estimates come from 6dFGS \citep{2012MNRAS.423.3430B,2017JCAP...05..015H,2017MNRAS.471..839A}, the ‘First Amendment’ set of SNe peculiar velocities \citep{2012MNRAS.420..447T}, 2MASS \citep{2011MNRAS.413.2906D}, GAMA \citep{2013MNRAS.436.3089B}, WiggleZ \citep{2012MNRAS.425..405B}, VVDS \citep{2008Natur.451..541G}, VIPERS \citep{2013A&A...557A..54D,2017A&A...604A..33P,2017A&A...607A..54H,2018A&A...610A..59M}, FastSound \citep{2016PASJ...68...38O}, SDSS+BOSS+eBOSS \citep{2021PhRvD.103h3533A}, DESI \citep{2026A&A...708A.219Q}, SDSS massE \citep{2024MNRAS.528.4922S}, Cosmicflows-4 \citep{2024MNRAS.531...84B}, and Pantheon+ \citep{2026MNRAS.550g1266S}.}
\label{fig:fig_6}
\end{figure*}

Fig. \ref{fig:fig_6} places our growth-rate constraints in the context of previous estimates from the literature and the \citet{2020A&A...641A...6P} $\Lambda$CDM prediction assuming general relativity. Both of our estimates are fully consistent with the GR prediction and with previous estimates at comparable redshifts. Under the tighter AP priors, our inferred growth rate shifts towards higher values, accompanied by a substantial reduction in uncertainty. The corresponding constraint is also higher than that of \citetalias{2021ApJ...920...13M}, despite the two analyses adopting the same geometric priors and fitting range, although the estimates remain statistically compatible. The comparison illustrates the sensitivity of the inferred growth rate to assumptions about the background geometry, with informative AP priors yielding tighter constraints, while marginalising over geometric distortions with broad priors provides a less restrictive estimate. Both results remain conditional on the adopted fiducial cosmological template. Both fits use the same clustering measurements under different assumptions about the background geometry and find no evidence for departures from general relativity within the adopted $\Lambda$CDM background cosmology.

\section{Conclusions}
\label{sec:conclusions}

We measured the monopole and quadrupole of the redshift-space 2PCF of a spectroscopic galaxy cluster sample drawn from the DESI Legacy Imaging Surveys to constrain the growth of structure and test consistency with general relativity. To enable this measurement, we developed and validated a method for reconstructing the angular and radial selection functions from the parent cluster catalogue to generate random catalogues. We also calibrated the effective linear halo bias using cluster abundance information, accounting for the spectroscopic selection, richness--mass relation, and its intrinsic scatter.

Our fiducial sample comprises \num{85818} clusters with richness $\lambda_{500}\geq13$ over $0.1\leq z \leq0.4$, corresponding to an effective redshift of $z_{\rm eff}\simeq0.28$. Using the one-loop perturbation-theory model implemented in \texttt{velocileptors}, we obtain $f\sigma_8=0.45\pm0.08$, corresponding to an uncertainty of approximately 18\%, after marginalising over the AP parameters with broad priors. Adopting tighter Gaussian priors on the AP parameters, following \citetalias{2021ApJ...920...13M}, and restricting the fitting range to exclude the BAO scales yields $f\sigma_8=0.52\pm0.03$, improving the precision to approximately 6\%. Both results are consistent with the predictions of general relativity within the adopted fiducial cosmology.

We assessed the robustness of these constraints against modelling and selection-function uncertainties, including the reconstruction of the survey footprint, foreground-dependent angular completeness, and the choice of clustering scales. Although the large-scale quadrupole exhibits a coherent negative residual relative to the best-fitting model, the full-covariance fit remains statistically acceptable, and excluding the affected measurements yields statistically compatible cosmological constraints. We also find that removing the abundance-derived prior on linear halo bias increases the uncertainty on $f\sigma_8$ by $25\%$, while the clustering data independently recover a bias consistent with the abundance prediction. These results demonstrate both the value of complementary cluster-abundance information and the importance of accurately characterising observational selection effects.

Forthcoming photometric cluster samples from Euclid \citep{2025A&A...697A...1E} and the Vera C. Rubin Observatory's Legacy Survey of Space and Time \citep{2012arXiv1211.0310L} will enable cluster-clustering studies with large, higher-redshift but photometrically identified samples. The proposed Wide-field Spectroscopic Telescope will substantially increase the number of spectroscopically-identified clusters both at low and high redshifts \citep{2024arXiv240305398M,2025arXiv251213221T}, reducing statistical uncertainties and enabling more precise measurements of redshift-space distortions. Extending spectroscopic cluster samples to higher redshifts will also make it possible to trace $f\sigma_8(z)$ over cosmic time and constrain the growth index through the redshift evolution of structure formation. Realising this potential will require increasingly accurate halo-bias calibrations, and comprehensive, publicly documented survey selection functions.

\begin{acknowledgements}
ET thanks Alan Heavens for helpful discussions. This research made use of ChatGPT (OpenAI; GPT-5.6 Sol) for assistance with language editing and coding. All AI-assisted material was reviewed, tested, and validated by the authors, who take full responsibility for the scientific content. This work was supported by STFC through Imperial College Astrophysics Consolidated Grant ST/W000989/1 in its original stages. This work made use of the marvin computing server funded by ANR grant ProGraceRay (ANR-23-CE31-0010) and hosted at LUX, Paris Observatory-PSL. 
\end{acknowledgements}

\bibliographystyle{aa}
\bibliography{main}

%
\begin{appendix}

\section{Mask reconstruction validation}
\label{ssec:mask_validation}

\begin{figure}
    \centering
    \includegraphics[width=0.5\textwidth]{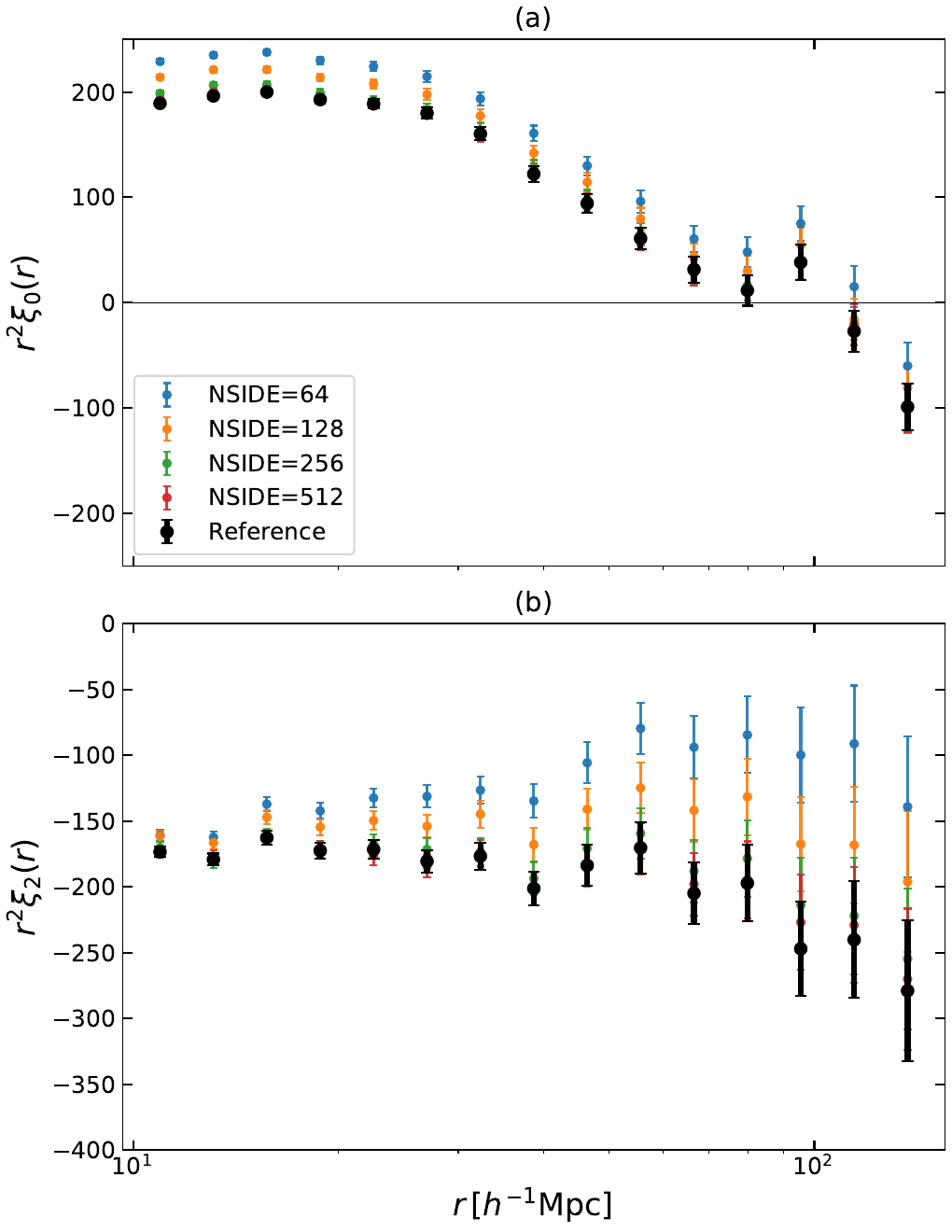}
    \caption{Measured 2PCF from the \texttt{PINOCCHIO} simulations with randoms constructed using the parent-sample method detailed in Sect. \ref{ssec:measurement}. $\mathrm{NSIDE}=512$ provides the best recovery of the clustering signal and stability of inferred cosmological parameters, while minimising the risk of imprinting cosmological structure onto the random catalogue.}
    \label{fig:fig_2}
\end{figure}

To assess the reliability of the mask reconstruction, we perform four complementary validation tests: (i) We compare the occupied-pixel mask with the large-scale footprint reconstructed from the parent catalogue using Delaunay triangulation. We assess whether the mask follows the reconstructed boundaries without introducing substantial regions outside the footprint. (ii) Using a third-order Lagrangian Perturbation Theory \texttt{PINOCCHIO} halo mock catalogue \citep{pinocchio,2017MNRAS.465.4658M} with the same configuration as our fiducial subsample, we measure the monopole and quadrupole with random catalogues generated from occupied HEALPix pixels at different angular resolutions. We compare these measurements against a reference obtained by independently resampling the angular positions and redshifts of the mock haloes, thereby testing the sensitivity of the recovered clustering signal to the mask reconstruction. (iii) We compare the observed fraction of occupied pixels with the expectation for a Poisson distribution of objects within the reconstructed footprint. We additionally perform a half-sample cross-validation, constructing the occupied-pixel mask from a random half of the parent catalogue and examining the spatial density of the remaining objects in retained and excluded pixels. Comparing the corresponding density ratios against Poisson expectations allows us to assess whether the mask preferentially traces fluctuations in the parent catalogue's surface density. (iv) Finally, we assess the stability of the inferred cosmological parameters with respect to the angular resolution of the reconstructed mask. We repeat the full clustering measurement and cosmological inference for different HEALPix resolutions, keeping the clusters in the selected sample, theoretical model, covariance prescription, and parameter priors fixed. We require our cosmological posteriors to stabilise with increasing NSIDE.

For the reference measurement of test (ii), we use one \texttt{PINOCCHIO} mock as the data catalogue and construct a random catalogue containing 20 times as many objects by sampling complete halo coordinates $(\mathrm{RA},\mathrm{Dec},z)$ without replacement from independent \texttt{PINOCCHIO} realisations, following the approach employed by \citet{2026A&A...713A.282E}. This avoids imprinting the particular angular clustering fluctuations of the data realisation onto the random catalogue. This approach yields 2PCF measurements consistent with theoretical model predictions \citep{2026A&A...713A.282E}. We then compare this reference measurement with those obtained using HEALPix-reconstructed random catalogues at different angular resolutions.

We find that $\mathrm{NSIDE}=512$ closely reproduces the reference \texttt{PINOCCHIO} monopole and quadrupole, provides good geometric agreement with the reconstructed footprint, and yields stable cosmological constraints. Furthermore, the observed pixel occupancy agrees with a Poisson expectation, while the half-sample cross-validation yields a density ratio between excluded and retained pixels of $1.003\pm0.012$, consistent with unity and indicating no detectable preferential selection of overdense regions. Coarser resolutions introduce systematic deviations, whereas increasing the resolution further provides no appreciable improvement and can introduce density-dependent selection effects. We therefore adopt $\mathrm{NSIDE}=512$ for the \citetalias{2024ApJS..272...39W} random catalogue.

\section{Effective bias validation}
\label{ssec:bias_validation}

We validate the abundance-based effective bias calibration by comparing the observed cluster abundance with predictions from the halo mass function combined with the fitted mass--proxy relation, examining the spectroscopic selection function, and assessing the sensitivity of the inferred bias to the redshift-abundance likelihood. We additionally investigate the redshift stability of the bias calibration through independent fits in three redshift intervals.

\begin{figure*}
    \centering
    \includegraphics[width=0.9\textwidth]{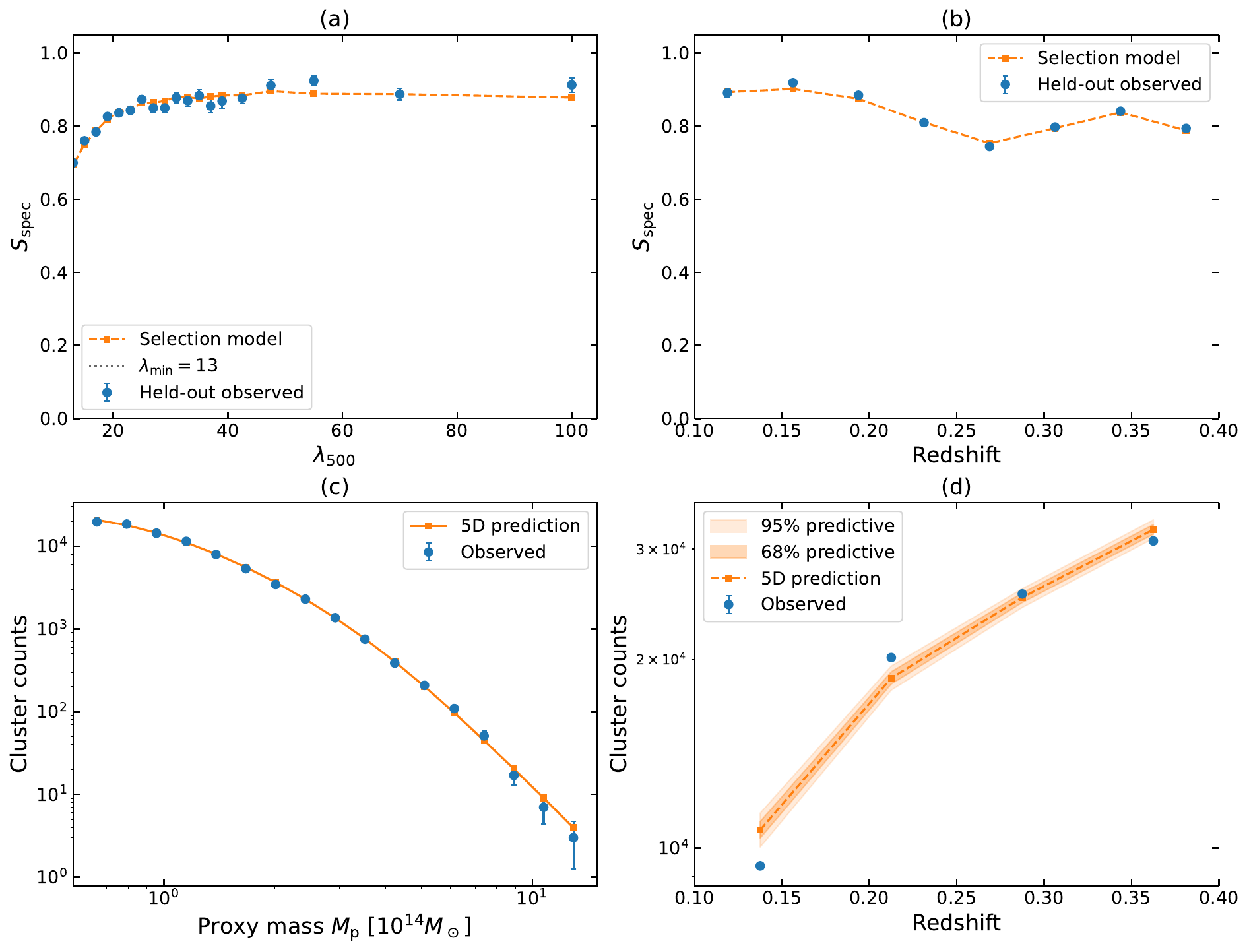}
    \caption{Validation of the abundance-based bias prediction for the \citetalias{2024ApJS..272...39W} cluster sample with $\lambda_{500}\geq13$ and $0.1\leq z \leq0.4$. Spectroscopic completeness as a function of (a) richness and (b) redshift, comparing the model predictions with the observed completeness in the independent 20\% hold-out sample. Observed and predicted cluster abundances as functions of (c) proxy mass and (d) redshift. The predictions in panel (c) use the posterior-median five-parameter mass--proxy relation, while panel (d) shows the median posterior prediction with 68\% and 95\% predictive intervals incorporating mass--proxy parameter uncertainties, Poisson noise, sample variance, and spectroscopic-selection uncertainty. The residual discrepancies in panel (d) have little impact on the effective bias prediction: reducing the contribution of the redshift-abundance likelihood by a factor of four changes the maximum-likelihood bias by $0.1\%$, while removing this contribution entirely changes it by $0.35\%$. The effective bias is weighted by the observed joint distribution of proxy mass and redshift rather than the model-predicted abundance.}
    \label{fig:fig_3}
\end{figure*}

Figure~\ref{fig:fig_3}a--b compares the predicted spectroscopic completeness with the observed fractions as functions of richness and redshift, respectively, using the independent 20\% hold-out sample. Panels (c) and (d) compare the observed and predicted cluster abundances as functions of proxy mass and redshift. The five-parameter abundance model, combining the halo mass function with the fitted mass--proxy relation, reproduces the observed proxy-mass distribution across the full sample, while the redshift-abundance comparison reveals residual discrepancies in some of the four redshift bins. To assess the redshift stability of the inferred bias, we also perform independent abundance fits in three redshift intervals and compare their effective bias predictions with those of the global five-parameter model. The resulting predictions are consistent at the 95\% credible level.

Finally, we assess the sensitivity of the effective bias prediction to discrepancies between the observed and modelled redshift abundance shown in Fig.~\ref{fig:fig_3}(d). We decompose the joint abundance likelihood into the conditional proxy-mass likelihood $P(M_{\rm p}\mid z)$ and the marginal redshift-abundance likelihood $P(N(z))$. We first reduce the contribution of the latter to the log-likelihood by a factor of four, finding a change of only $0.1\%$ in the maximum-likelihood effective bias. We then exclude this contribution entirely, fitting only the conditional proxy-mass distribution within each redshift interval. The resulting effective bias differs from the fiducial joint-likelihood prediction by $0.35\%$. Both tests indicate that the bias calibration is relatively insensitive to discrepancies in the modelled redshift abundance.

\end{appendix}

\end{document}